\documentclass[letterpaper,hyper]{JHEP3}
\usepackage{epsfig}
\usepackage{amsmath, amsfonts,euscript,bbm}
\usepackage{ifthen}
\usepackage{graphicx}

\newcommand{\be}{\begin{eqnarray}}

\newcommand{\ee}{\end{eqnarray} }

\newcommand{\Mpl}{M_{\rm pl}}
\newcommand{\ex}[1]{\langle #1 \rangle}

\newcommand{\at}{\tilde a}

\newcommand{\csv}{c_{sv}^{2}}
\newcommand{\cbv}{c_{bv}^{2}}
\newcommand{\cs}{c_{s}^{2}}
\newcommand{\cH}{\mathcal{H}}

\newcommand{\dl}{\delta_{l}}

\newcommand{\dej}{\Delta J}
\newcommand{\cv}{c_{v}^{2}}
\newcommand{\dsq}{\Delta^{2}}

\title{On the Renormalization of the Effective Field Theory of Large Scale Structures}

\author{Enrico Pajer${}^{1}$ and Matias Zaldarriaga${}^{2}$, \\
$^1$ Department of Physics, Princeton University, Princeton, NJ 08544, USA\\
$^2$ Institute for Advanced Study, Princeton, NJ 08544, USA}

\date{}

\abstract {Standard perturbation theory (SPT) for large-scale matter inhomogeneities is unsatisfactory for at least three reasons: there is no clear expansion parameter since the density contrast is \textit{not} small on all scales; it does not fully account for deviations at large scales from a perfect pressureless fluid induced by short-scale non-linearities; for generic initial conditions, loop corrections are UV-divergent, making predictions cutoff dependent and hence unphysical. 

The Effective Field Theory of Large Scale Structures successfully addresses all three issues. Here we focus on the third one and show explicitly that the terms induced by integrating out short scales, neglected in SPT, have exactly the right scale dependence to cancel all UV-divergences at one loop, and this should hold at all loops. A particularly clear example is an Einstein deSitter universe with no-scale initial conditions $P_{in}\sim k^{n}$. After renormalizing the theory, we use self-similarity to derive a very simple result for the final power spectrum for any $n$, excluding two-loop corrections and higher. We show how the relative importance of different corrections depend on $n$. For $n\sim-1.5$, relevant for our universe, pressure and dissipative corrections are more important than the two-loop corrections.}

\begin{document}


 
\section{Introduction}

On very large scales, inhomogeneities in our universe are small, of the order $10^{-5}$. On small scales, due to gravitational collapse, clumps of matter form. Numerical methods such as N-body simulations can tackle the full non-linear problem and therefore have been widely employed to describe all scales. In recent years, with the development of the Baryon Acoustic Oscillation (BAO) technique as a probe of the expansion history of the Universe, interest has grown in the so-called mildly non-linear regime of structure formation. These are scales where non-linearities are small and thus amenable to perturbative methods but still non-negligible to interpret precision measurements. Numerical simulations can be used to study this regime, although to probe cosmological models whose parameter space has a large dimensionality requires significant computational resources.  In addition,  numerical simulations may tend to make the underlying physics opaque, or at least one is much more satisfied when one can achieve a relatively accurate analytic understanding of what the simulations show. This is why, despite the impressive development in the computational speed, there is still a large efforts to improve analytical methods and thus also improve our understanding of the evolution of structures. Most analytical methods are based on a perturbative expansion in small inhomogeneities and therefore are valid only in between the Hubble scale $H$ and some non-linear scale $k_{NL}\gg a H$, with $a$ the FLRW scale factor. Much work, including the present one, has been devoted to pushing the accuracy of the analytical methods closer to the non-linear scale and understanding its physical origin.

In the standard (Euclidean) perturbation theory (SPT) (see \cite{BCGS} are references therein)  the final results, e.g.~for the density power spectrum, are organized in a loop expansion where the small parameter is the initial power spectrum of inhomogeneities. There are at least two conceptual issues worth emphasizing: first inhomogeneities are large at small scales, which makes it hard to rigorously define the expansion parameter. Furthermore at sufficiently small scales multiple streams develop so the fluid equations typically used become invalid. Second, for generic initial conditions, loop integrals are typically divergent, but no counterterms appear in the theory with the appropriate scaling to cancel them. This means that the predictions depend on the cutoff and are hence unphysical. The sensitivity to the cutoff also points to the fact that the answer depends on small scales where the physics in not being modeled properly. 

Both issues are successfully addressed within the Effective Field Theory (EFT) of Large Scale Structures (LLS) \cite{BNSZ,CHS,H}, henceforth EFToLSS. The main idea is the usual underlaying principle of EFT. Integrate out the short-scale modes and systematically incorporate their effect on the large-scale dynamics (see also \cite{P} for a similar approach). Starting with a configuration of collisionless dark matter particles in the Newtonian limit in an expanding universe, one derives a system of fluid equations. The effects of the short scales is to induce a non-vanishing \textit{speed of sound}, \textit{dissipative corrections} and a \textit{stochastic noise} on top of the perfect fluid that one would have obtained if these effects had been neglected. On the one side, this is remarkable if one remembers that we started with collisionless particles. On the other this is precisely what we should have expected based on the EFT philosophy: all terms compatible with the symmetries of the problem should be present. These additional terms encode the effects from the small scales whose dynamics is not captured by the EFT on the large scales which are now correctly modeled. 

The goal of this paper is to clarify how the approach of the EFToLSS solves the second problem mentioned above, namely the fact that SPT predictions are cutoff dependent and hence unphysical. Building on the results of \cite{CHS,H}, we show in detail that the effective terms induced by integrating out the short modes have exactly the right scale and time dependence to cancel the divergent terms. This result is general and makes a strong case that the EFToLSS, rather than SPT, is the theoretically consistent way to perform Euclidean perturbation theory. In order to show this in the clearest possible way, starting in section \ref{1}, we focus on an Einstein de Sitter (EdS) cosmology containing only non-relativistic matter. This has a few advantages: first EdS is a good description of our universe after matter radiation equality with the largest deviations arising at late times when Dark Energy starts dominating; second, thanks to the simple form of the growth factor in perturbation theory almost all results are amenable to analytical computations; finally, provided the right initial conditions are specified, namely a no-scale initial power spectrum $P_{\delta}\sim k^{n}$, the system enjoys a \textit{self-similar evolution}, which ensures that analytical solutions can be constructed that are valid at all times. 

We discuss the renormalization of the theory in section \ref{divergences}. We employ both a hard cutoff, which breaks self-similarity, and dimensional regularization which preserves it. The final results agree with each other and satisfy self-similar scaling as expected. For EdS we write down explicit formulae for the final non-linear density power spectrum (without including two-loop terms and higher)  including the leading effective corrections for any $n$. Our final result is given in \eqref{finres}. We also highlight the relative importance of the different terms as the non-linear scale is approached (see figure \ref{ns}). Our expression \eqref{finres} contains zero, one or two free parameters (excluding two-loop corrections or higher), depending on $n$. In the specific case $n\sim-1.5$, which is appropriate for our universe close to the present non-linear scale, only one parameter appears up to two-loop correction, namely the one associated with the speed of sound (whose phenomenological effects were studied in \cite{CHS,H}). We present some preliminary comparison with results from published numerical simulations in EdS for $n=-1.5$ and $n=-1$ in section \ref{s:num} and show that there is good agreement, but clearly more work is required to make a more convincing comparison.  We conclude in section \ref{end} with a discussion and ideas for future research.

 
\section{Smoothed perturbation theory}\label{1}

Accounting for the effects of integrating out small scales, the equations of motion for the density contrast $\delta\equiv \rho/\bar \rho-1$ with $\bar \rho$ the average density, the velocity $v^{i}$ and the gravitational potential $\phi$ in the Newtonian limit (neglecting vorticity, which decays at linear order) around an FLRW universe are \cite{BNSZ,CHS}
\be\label{all}
\partial_{\tau} \dl+\partial_{i} \left[ \left(1+\dl\right) v_{l}^{i} \right]&=&0\,,\\
\partial_{\tau}v^{i}_{l}+\cH v^{i}_{l}+\partial_{i} \phi +v_{l}^{k}\partial_{k} v_{l}^{i}&=&-\cs \partial^{i}\dl+\frac{3}{4} \frac{\csv}{\cH} \partial^{2}v_{l}^{i}+\frac{4\cbv+\csv}{4\cH} \partial^{i}\partial_{j}v_{l}^{j}-\Delta J^{i}\dots\,,\label{NS}\\
\partial_{i}\partial_{i} \phi_l&=& \frac{a^{2}}{2} \bar \rho \dl= \frac{3}{2} \cH^{2} \Omega_{m} \dl = \frac{3}{2} \frac{\cH_{0}^{2}\Omega_{m,0}}{a}\dl\,,
\ee
where $\tau$ is conformal time, $\cH\equiv\partial_{\tau}a/a=aH$ with $H\equiv \partial_{t} a/a$ and $\Omega_{m}(\tau)\equiv\bar \rho/3H^{2}\Mpl^{2}$ with $\Omega_{0}=\Omega(a_{0}=1)$. Some additional comments are in order. The subscript $l$ indicates that the corresponding field has no short-wavelength fluctuations, i.e.~it has been smoothed on a certain scale $\Lambda$. These equations can be solved perturbatively when $\Lambda<k_{NL}$ (we will define $k_{NL}$ in \eqref{kNL}), so that the smoothed long-scale perturbations are never large. On the right hand side of \eqref{NS} one finds the corrections induced by the short scales. This is a double expansion in perturbations and derivatives acting on them. At zeroth and linear order in perturbations to lowest order in derivatives we have four terms. Each one of them comes with a infinite series of higher derivative corrections which we include in the ellipsis. $\cs,\cbv$ and $\csv$ are real time-dependent functions that physically correspond to the speed of sound, bulk and sheer viscosity respectively. $\dej^{i}$ is a time and scale dependent random variable representing stochastic noise induced by the fact that any specific short-scale configuration deviates somewhat from the ensemble average. The values (and distribution function in the case of $\dej^{i}$) of these parameters are not predicted within the EFToLSS approach. Rather, they can be computed in the microscopic theory or, when this is too hard, they should be extracted by comparison with observations or numerical simulations.

 
\subsection{Loop corrections in an Einstein de Sitter universe}

One of the main goals of the present paper is to show that the new terms induced by integrating out the short scales have exactly the right scale dependence to cancel the UV divergences arising in perturbation theory. Because the EFToLSS is derived by including all possible terms compatible with the symmetries of the problem (homogeneity and isotropy of the background, conservation of mass. etc.) this should be true in any cosmology at all loops. For concreteness we consider an EdS universe because the formulae simplify considerably and because of its relevance for our universe. We also neglect two-loop and higher terms, but expect that our results generalize straightforwardly to higher orders.

Consider a flat universe filled only with non-relativistic matter, a.k.a.~an Einstein-deSitter (EdS) universe. Then $\Omega_{m}=1$ and one finds
\be
\cH=\frac{2}{\tau}\,,\quad a= \left(\frac{\tau}{\tau_{0}}\right)^{2}\,,\quad \cH_{0}=\cH \sqrt{a}\,,\quad \cH'= -\frac{\cH}{2a}\,.
\ee
Neglecting for the moment the effective terms in the right-hand side of \eqref{NS}, the system reduces to the usual SPT equations whose perturbative solutions are well known (see e.g.~\cite{BCGS}). Here we simply quote the one-loop result for the correction to the power spectrum
\be
P(a,k)= P_{lin}(k,a)+P_{22}(k,a)+  P_{13}(k,a)+P_{\cs}(k,a) + P_{J}(k,a)\,,
\ee
with
\be
P_{lin}&=&a^{2}P_{in}\,,\\
P_{13}(k)&=&\frac{a^{4}}{ 252}\frac{k^3}{ 4\pi^2}P_{in}(k)\int dr\,P_{in}(k\,r)\nonumber\\
&&\quad\left({12\over r^2}-158+100r^2-42r^4+{3\over r^3}(r^2-1)^3(7r^2+2)\ln\left|{1+r\over1-r}\right|\right)\label{P13}\,,\\
P_{22}(k)&=&{a^{4}\over 98}{k^3\over 4\pi^2}\int dr\int dx\,P_{in}(k\,r)P_{in}(k\sqrt{1+r^2-2rx})    {(3r+7x-10rx^2)^2\over(1+r^2-2rx)^2}\,.\label{P22}
\ee

Before discussing the new lowest order corrections induced by the short modes, namely $P_{\cs}$ and $P_{J}$, let us pause and comment on the one-loop terms $P_{13,22}$. These can be both UV and IR divergent depending on the initial power spectrum under consideration. For no-scale initial conditions $P_{in}=A k^{n}$ with $A$ a constant, the conditions for the occurrence of divergences are summarized in table \ref{t:div}. For the moment we will ignore these divergencies but will discuss them in detail in section \ref{divergences}.

To compute these corrections we notice that at linear order in perturbations one can find a second order differential equation for $\dl$ (valid in any FLRW cosmology)
\be\label{delta:eq}
-a^{2}\cH^{2} \partial_{a}^{2}\dl-a \left(2\cH^{2}+a\cH\cH_{,a}\right) \partial_{a}\dl+\frac{3}{2}\cH^{2}\Omega_{m}\dl&=&-\cs \partial^{2} \dl- \cv a \partial^{2}\partial_{a} \dl-J\,,
\ee
where we used the scale factor $a$ as time variable and defined $\cv \equiv \cbv+\csv$ and $J\equiv \partial_{i}\Delta J^{i}$. Let us treat the terms on the right-hand side perturbatively. Since the left-hand side of equation has no spatial derivatives, the zeroth-order solutions is given in the form 
\be
\dl=\delta_{i}(k) \left[C_{+}D_{+}(a)+C_{-}D_{-}(a)\right]\,,
\ee
with $C_{\pm}$ two integration constants and $\delta_{i}(k)$ a stochastic variable fixed by some boundary condition. The growing and decaying solutions are $D_{+}=a$ and $D_{-}=a^{-3/2}$ and in the following we will discard the latter. The retarded Green's function in EdS (we discuss the d-dimensional generalization of this result in appendix \ref{d}) takes the form
\be\label{G}
G(a,\at)=\theta(a-\tilde a) \frac{2}{5} \cH_{0}^{-2} \left[  \left(\frac{\at}{a}\right)^{3/2}-\frac{a}{\at}\right]\,.
\ee
Notice that
\be\label{usef}
\int d\at G(a,\at) \at^{n}=-\frac{ a^{n+1}}{\cH_{0}^{2}n \left(n+5/2\right)}\,,
\ee
which explains the well know structure of SPT in EdS $\delta_{n}\propto a^{n}$.

 
\subsection{Self-similar solutions}\label{sss}


In an Einstein de-Sitter cosmology, the equations of motion posses a scaling symmetry. 
For simplicity let us consider the case when matter can be treated as a cold fluid.  Although this is not the case on small scales and we should be studying the dynamics of collisionless particles, the equations for a collection of non-interacting particles has the same symmetry we now discuss. The equations are just those in \eqref{NS} neglecting for the moment the effective terms in the right hand side. 
\be\label{full}
\partial_{\tau} \delta+\partial_{i} \left[ \left(1+\delta\right) v^{i} \right]&=&0\,,\\
\partial_{\tau}v^{i}+\cH v^{i}+\partial_{i} \phi +v^{k}\partial_{k} v^{i}&=&0\,,\\
\partial_{i}\partial_{i} \phi&=&  \frac{3}{2} \cH^{2} \delta ,
\ee
where $\cH = 2/\tau$. One can take a solution of these equations $[\delta(x,\tau), v^{i}(x,\tau),\phi(x,\tau)]$ and  scale it to get another one: 
\be
\tilde \delta(x,\tau)&=&\delta(\lambda_x x, \lambda_\tau\tau)\,, \label{s1}\\
\tilde v^i(x,\tau)&=&{\lambda_\tau \over \lambda_x} v^i(\lambda_x x, \lambda_\tau\tau)\,,\\
\tilde \phi(x,\tau)&=&\left({\lambda_\tau \over \lambda_x}\right)^2\phi(\lambda_x x, \lambda_\tau\tau)\,.\label{s3}
\ee 
That this is a solution can be checked by simply replacing it in eq. (\ref{full}). Although in this way we obtain a new solution for any choice of $\lambda_x$ and $\lambda_\tau$, not all the choices correspond to a realization of the same cosmology. By this we mean that not all solutions have initial conditions sampled from the same power spectrum.  By making a specific choice of the relative $x$ and $\tau$ scalings one can get a family of solutions such that for any $\lambda$ they are just different realizations with the \textit{same} initial statistical properties. 

Consider $\Delta^2(k,\tau)$,
\be
\dsq(k,\tau)\equiv \frac{k^{3}P(k,\tau)}{2\pi^{2}}\,, \label{D} \quad {\rm with} \quad \ex{\delta(\vec k)\delta(\vec k')}=(2\pi)^{3}\delta^{3}(\vec k-\vec k') P(k,\tau)\,.
\ee
After scaling we have:
\be
\tilde \Delta^2(k,\tau)= \Delta^2(k/\lambda_x,\lambda_\tau \tau).
\ee

We will first study this relation at very early times such that all quantities can be computed in linear theory. We will assume that $[\delta(x,\tau),v^{i}(x,\tau),\phi(x,\tau)]$ were sampled from a Gaussian distribution. The time dependence of the scaled solutions is the same as the  original one so both are in the correct growing mode solution. The only remaining issue is to make sure that the scaled solutions have the same power spectrum at early times. Assuming power law initial conditions for $\delta$ with index $n$, i.e.~$\Delta^2_{initial}(k,\tau)\propto \tau^4 k^{n+3}$, we get
\be
\tilde \Delta^2_{initial}(k,\tau)\propto { \lambda_\tau^4 \over \lambda_x^{n+3}}  \Delta_{initial}^2(k,\tau).
\ee

Thus if we pick 
\be\label{lambdas}
\lambda_x=\lambda_\tau^{4\over n+3},
\ee
then the different rescaled solutions are just different samples of the same initial power spectrum. If this is so, then at any given time the power spectra of all the scaled solutions (which is an ensamble averaged quantity)  should be the same as the original un-scaled solution. Thus, 
\be
\Delta^2(k,\tau) =\tilde \Delta^2(k,\tau) = \Delta^2(k/\lambda^{4\over n+3},\lambda \tau).
\ee
This is satisfied if 
\be
\Delta^2(k,\tau) =\Delta^2(k/k_{NL}),
\ee
where 
\be\label{kNL}
k_{NL}^{3+n}\equiv \frac{2 \pi^{2}}{A a^{2}}\propto \tau^{-4},
\ee
 so that $k/k_{NL}$ is invariant under the scaling \eqref{lambdas}. With this definition,  $\Delta_{lin}(k_{NL})\equiv 1$.

 Perhaps this is just a complicated way of saying that the only scale in the problem is the non-linear scale and thus the answer should just be expressed in terms of it. Of course in practical situations, such as for example when running a numerical simulation with finite computer resources, one is forced to introduce additional scales such as the size of the box or the inter-particle separation. As long as these scales are sufficiently removed from the non-linear scale their effects should be small. This has been checked in practice for a wide range of $n$'s. Convergence is progressively more difficult to achieve the closer the spectrum is to $n=-3$ at which point all scales become non-linear at the same time. 


Given that the full solution satisfies the scaling symmetry, both at long and short scales, the new terms appearing in the smoothed equations of the EFToLSS \eqref{NS} and  (\ref{delta:eq}) should also satisfy it, since they arise from integrating out the short scales. This implies that each term should scale according to its dimensions. For example: 
\be
 \cs,c_{bv}^{2},c_{sv}^{2}&\propto&  \frac{\lambda_x^2}{\lambda_\tau^2} \,.
\ee
Since these terms represent averages over the short wavelengths, the scaling symmetry should only be respected if we consider the combined $x-\tau$ rescaling that leaves the initial spectrum fixed, namely \eqref{lambdas}. We can then use (\ref{lambdas}) and \eqref{delta:eq} to deduce the appropriate time dependence of the sound speed coefficients, 
\be
 \cs,c_{bv}^{2},c_{sv}^{2}\propto \tau^{\frac{2-2n}{n+3}}\,.\label{cssc}
\ee
This scaling of $c_{s}^{2}$ can be derived as well by requiring that the Jeans scale is a constant fraction of the non-linear scale, since again there should be only one independent scale in the problem.

The properties of $J$ in (\ref{delta:eq})  can be similarly obtained. In this case the power spectrum of $J$ should scale in such a way as to preserve the scaling symmetry under the combined $x-\tau$ scaling. The only additional complication now is that $J$, unlike $\cs$, is spacetime dependent, and hence can contain and unknown function of $k/k_{NL}$, which, being invariant under the combined $x-\tau$ scaling, can not be fixed with the arguments we have used so far. In particular, by inspection of \eqref{NS}, one finds
\be
\Delta^2_J(k,\tau) =\tau^{-4} f(k/k_{NL}),
\ee
where $f$ is some arbitrary function. In the limit of $k\ll k_{NL}$ the $k$ dependence can be constrained following arguments in \cite{Peebles}. Mass and momentum conservation imply that on large scales the corrections to the density power spectrum from these terms scale as $k^4$ rather than $k^0$ or equivalently:
\be\label{jsca}
\Delta^2_J(k,\tau) \propto  \tau^{-4} (k/k_{NL})^7 \ \ \ \ \  (k\ll k_{NL}). 
\ee


\subsection{The self-similar power spectrum}\label{pspec}

Now that we have the scaling of the effective coefficients, we can derive a formula for the power spectrum. For this section we will ignore the issue of divergences, postponing a thorough discussion to the next section. Because of self-similarity, we expect that the final $\Delta_{tot}^{2}$ (and not $P_{tot}$) should be a function of just $k/k_{NL}$. For this to be true the $k$ dependence of the various terms will have to combine with the $\tau$ dependence in a precise way. This will provide us with a cross check of the scalings computed in the previous section.

Let us start considering the corrections coming from $\cs$ and $\cv$. Their leading order contribution to $\delta$ (from now on we drop the label $l$ for large scale perturbations, leaving it implied that all the fields we will be discussing have been smoothed above a certain scale $\Lambda<k_{NL}$) perturbations is given by
\be
\delta_{\cs}=\int da G(a,\tilde a)  k^{2} \left(\cs+\cv a\partial_{a}\right)\delta_{1}\,,
\ee
where $\delta_{1}$ is the linear order perfect fluid solution, $G$ the Green's function in \eqref{G} and in EdS $a=(\tau/\tau_{0})^{2}$. The linear contribution of this term to the power spectrum is hence found to be
\be\label{csca}
\Delta_{\cs}^{2}\propto k^{5} \left(\cs+\cv \right) P_{in}(k) \,\tau^{4+2}\,,
\ee
where the extra $\tau^{2}$ comes from the Green's function. For self-similar initial conditions $P_{in}=A k^{n}$, the exponent of $k$ is $5+n$, which perfectly combines with the exponent of $\tau$, once the scaling \eqref{cssc} has been taken into account, to give $ \left(k/k_{NL}\right)^{5+n}$, up to higher derivative corrections.

For the noise term things work out analogously. At linear order one has
\be
\delta_{J}=- \int da G(a,\tilde a)  J\,,
\ee 
and hence
\be\label{jsca2}
\Delta_{noise}^{2}\propto \Delta_{J}^{2}(k) \tau^{4}\,,
\ee
where the $\tau^{4}\sim a^{2}$ comes from the Green's function. This again combines perfectly to give $ \left(k/k_{NL}\right)^{7}$, up to higher derivative corrections.
Finally, let us discuss higher loop corrections. Since loops count the exponent of the linear power spectrum, one expect $\Delta_{N-loop}^{2}\propto \left(k/k_{NL}\right)^{(n+3)(N+1)}$.

In summary, the self-similar power spectrum is given by
\be
\Delta^{2}_{tot}&=&\Delta^{2}_{lin}+\sum_{N}\Delta^{2}_{N-loop}+\Delta^{2}_{\cs}+\Delta^{2}_{noise}\,,\\
\Delta^{2}_{lin}&=& \left(\frac{k}{k_{NL}}\right)^{n+3}\,,\\
\Delta^{2}_{N-loop}&\propto& \left(\frac{k}{k_{NL}}\right)^{(n+3)(N+1)}\,,\\
\Delta^{2}_{\cs}&\propto&  \left(\frac{k}{k_{NL}}\right)^{n+5}+{\rm higher\, derivatives}\\
\Delta^{2}_{noise}&\propto&  \left(\frac{k}{k_{NL}}\right)^{7}+{\rm higher\, derivatives}\,.
\ee
and it is just be a function of $k/k_{NL}$ as anticipated.

\section{Cancellation of divergences and final power spectrum}\label{divergences}


It is time to confront the issue we neglected so far.  As we stated in table \ref{t:div}, the total one-loop contribution is UV divergent for $n\geq-1$ and IR divergent for $n\leq -3$. For $n\geq -3$ both $P_{13}$ and $P_{22}$ are IR divergent but their divergences cancel each other exactly. This is expected to happen at all orders and was discussed more in details in \cite{IR}. In this paper we simply focus on $n\geq-3$ and safely ignore the issue of IR divergences\footnote{For the case $n=-1$ there are IR divergences that leak into the computation of $P_{13,22}$ when performed using dimensional regularization. As all other IR divergences for $n\geq -3$ also these log-divergences cancel exactly once we add up the two one-loop contributions.\label{IRdiv}}. UV divergences instead appear naively more problematic (see e.g.~\cite{FS}). First, in order to get finite results one needs to regularize the theory. The details of how things work out depend on the regularization procedure and we will discuss cut-off regularization and dimensional regularization separately. 

\begin{figure}
\centering
\includegraphics[width=.8\textwidth]{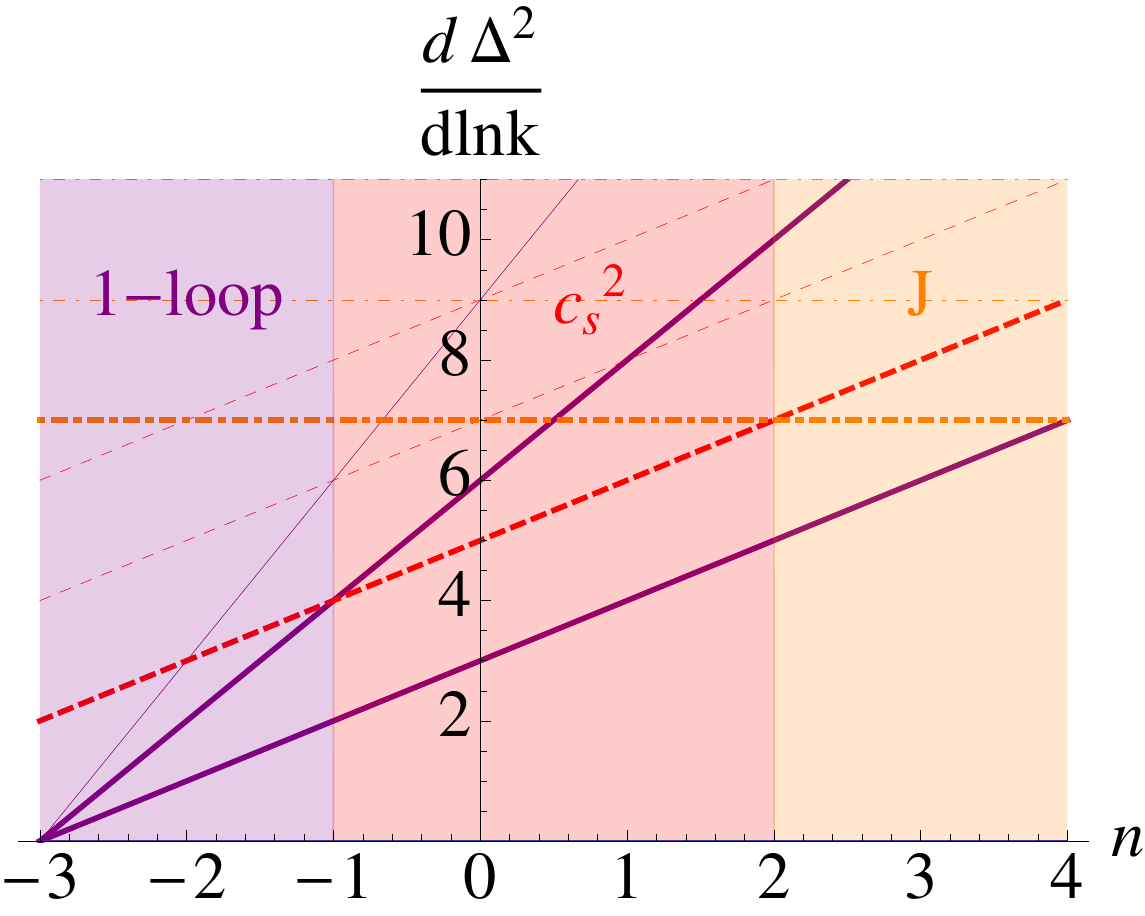}
\caption{We show the exponent of $k/k_{NL}$ for various contributions to the power spectrum as function of the power in the initial conditions $P_{lin}\propto k^{n}$. The full orange lines are zero, one and two loop contributions from SPT (from bottom to top); the black dot-dashed line is the term $\Delta_{\cs}$ while the dotted red line refers to $\Delta_{J}$. The three background regions labeled 1-loop, $\cs$ and $J$ indicate which is the most important correction to $\Delta_{lin}$ at that value of $n$. For our universe, near the non-linear scale $n\sim-1.5$ (eg. \cite{Smith:2002dz}). \label{ns}}
\end{figure}

The final result without including two-loop terms and higher is: 
\be
\dsq&=& \left(\frac{k}{k_{NL}}\right)^{3+n} \left\{1+\left(\frac{k}{k_{NL}}\right)^{3+n} \left[\alpha(n) +\tilde \alpha(n)\ln \left(\frac{k}{k_{NL}}\right)\right]\right\}\label{finres}\\
&&\quad +\beta(n) \left(\frac{k}{k_{NL}}\right)^{5+n} \left[1+{\rm higher\,deriv.}\right] +\gamma(n) \left(\frac{k}{k_{NL}}\right)^{7} \left[1+{\rm higher\,deriv.}\right]\,.\nonumber
\ee
where $\beta$ and $\gamma$ are fitting coefficients, $\alpha$ and $\tilde \alpha$ can be computed in SPT (see table \ref{t}) and higher derivatives come with even positive powers of $k/k_{NL}$. We decompose\footnote{This is related to the $\alpha_{\delta}$ in \cite{BCGS} by $\alpha_{\rm here}=\alpha_{\delta}^{BCGS} \Gamma \left[(3+n)/2\right] \frac{1}{2}$. For this conversion it is important to notice that eq.~(168) of \cite{BCGS} is \textit{not} valid in general, but only for $n=-2$. The general relation is instead $(k_{NL}R_{0})^{3+n}=\Gamma \left[(3+n)/2\right] \frac{1}{2}$} $\alpha(n)=\alpha_{22}(n)+\alpha_{13}(n)$ and the same for $\tilde \alpha$, with 
\be
\alpha_{13,22}+\tilde \alpha_{13,22} \ln \left(\frac{k}{k_{NL}}\right)\equiv \frac{P_{13,22}}{k^{3+2n}}\frac{2\pi^{2}}{ A^{2}a^{4}}\,,
\ee 
and give the explicit values in table \ref{t}. Note that, because we did not compute two-loop and higher terms, only terms that scale with a lower power of $k/k_{NL}$ than the two-loop term, namely $(k/k_{NL})^{3(n+3)}$, should be included in a consistent calculation. Figure \ref{ns} shows the $k/k_{NL}$ scaling of the various terms as a function of $n$. For example in the absence of two loop terms  it is only consistent to include noise corrections for $n>-0.5$ and the sound speed terms for $n> -2$. For $n$ relevant to the non-linear scale in our late universe, i.e.~$n\simeq -1.5$, the sound speed corrections are more important than the two-loop terms but the noise corrections are even less important than three loop terms. Finally notice that for those $n$'s for which log's appear, the finite loop term has the same $k$-dependence as either the noise or the speed of sound corrections (we will discuss this more in the following). In these cases $\alpha$ is degenerate with either $\gamma$ or $\beta$, which are fitting parameters, and hence its value is irrelevant and we omit it.

\begin{table}
\centering
\begin{tabular}{|c|c|c|}  \hline
               & UV div         & IR div    \\ \hline\hline
$P_{13} $&$  n \ge -1  $&$ n\le -1$ \\ \hline
$P_{22}$ & $ n \ge 1/2 $& $ n\le -1 $ \\ \hline
$P_{total}$ & $ n \ge -1 $& $ n\le -3 $ \\ \hline
\end{tabular}
\caption{The table shows for which $n$ UV or IR divergences arise at one-loop order.\label{t:div}}
\end{table}


\subsection{Cut-off regularization}\label{cutoff}

Let us now discuss how to regularize the theory by introducing a UV cutoff on the momenta. This choice is very natural in the EFToLSS approach since the fields have been smoothed over scales of order $\Lambda$ from the beginning. Then all loop integrands, which include positive powers of the power spectrum, go quickly to zero when the argument of any of the power spectra is much larger than $\Lambda$. This makes the loop integrals finite and cutoff dependent. This cutoff is time independent in the derivation of EFToLSS and hence breaks self-similarity explicitly. This is  not a problem per se, since as we will see the final results after renormalization are cutoff independent and hence preserve self-similarity as expected. In fact a regulator that does not respect the symmetries of a theory is not uncommon in (quantum) field theory; it is a consistent procedure but it makes the steps preceding the final result less transparent. In the next subsection we consider dimensional regularization, which preserves self-similarity throughout. 

The one-loop divergences can be seen by expanding \eqref{P13} and \eqref{P22} for large $q$. Up to numerical coefficients one finds
\be
P_{13}&\simeq&a^{4}k^{2}P_{11}(k) \int^{\Lambda} dq \,P_{11}(q) \,,\label{P13div}\\
P_{22}&\simeq&a^{4}k^{4} \int^{\Lambda} \frac{dq }{q^{2}}P_{11}(q)^{2} \,,\label{P22div}
\ee
from which the results for the UV divergences in table \ref{t:div} can be easily extracted. These two terms have precisely the same $k$ dependence as $P_{\cs}$ in \eqref{csca} (as already discussed in \cite{CHS,H}) and $P_{noise}$ in \eqref{jsca} and \eqref{jsca2}, respectively ($\Delta^{2}$ was defined in \eqref{D}). This means that any $\Lambda$ dependence arising in the two one-loop integrals above can be absorbed by a counterterm in the form of a specific $\cs(\Lambda)$ or $\Delta_{J}^{2}(\Lambda)$. This is a crucial point so let us stress it again. The additional terms present in the EFToLSS, which arise from integrating out the short scale degrees of freedom and which are neglected in SPT, are precisely the terms needed to cancel the cutoff dependence and hence make physically meaningful predictions (this was recognized already in \cite{BNSZ,CHS,H}).

It can be easily checked that the counterterms needed to cancel the divergences do \textit{not} have the correct time dependence required by self-similarity, but this is to be expected. Since the cutoff regularization breaks self-similarity explicitly by introducing another independent scale $\Lambda$, also the counterterms violate it. The important point is that, after the counterterms have canceled the divergences, the final results respect self-similarity.

After renormalization the only terms that survive are the finite cutoff independent parts of the loop-integrals, which can be computed using the expressions \eqref{P13} and \eqref{P22} (or \eqref{FS1} and \eqref{FS2} using dimensional regularization), and the finite parts of $\cs$ and $\Delta_{J}^{2}$, which scale as discussed in subsection \ref{pspec}. Since the details of the computation have been presented elsewhere \cite{FS} and are lengthy and not particularly illuminating, we do not present them here, but we attach to our publication a mathematica file \cite{file} that computes, as an example, the finite part of $P_{13}$ for $-3\leq n\leq 4$ ($P_{22}$ is slightly more difficult because of the angular integral, but can be computed with some work \cite{FS}).

\begin{centering}

\begin{table}\label{t}
\hspace{-.5cm}
\begin{tabular}{|c|c|c|c |c|c|c|c |c|c|c|c|}  \hline
n                            & -2                                &	-3/2 	                            &  -1                                        & -1/2        &                                 0                                   & 1/2                                   & 1                                       & 3/2     & 2                                                                           &   5/2                                          & 3 \\ \hline\hline
$\alpha_{13} $ & $  \frac{5 \pi^{2}}{112}   $&$ \frac{992 \pi}{6,615}  $&$  \dots $&$  -\frac{416 \pi}{8,085}	$&$  -\frac{\pi^{2}}{336}    $&$   \dots  $&$ \dots $&$   \dots $&$ -\frac{\pi^{2}}{168} $&$ \dots $&$ \dots  $\\ \hline

$\alpha_{22} $ & $  \frac{75 \pi^{2}}{784} $&$                   -0.232       $&$  \dots $&$              .698                 $&$ \frac{29 \pi^{2}}{784} $&$  \dots   $&$       \dots   $&$   \dots $&$ \frac{\pi^{2}}{392} $&$ \dots $&$ \dots $\\ \hline
$\tilde\alpha_{13} $ &$  0  $&$ 0 	$&$ \frac{61}{315}  $&$ 0 $&$ 0 $&$ 0 $&$ -\frac{4}{105} $&$ 0 $&$ 0 $&$ 0 $&$ \frac{20 }{1,323} $\\ \hline
$\tilde\alpha_{22}$ &$ 0  $&$ 0 $&$  0  $&$ 0 $&$ 0 $&$ -\frac{9}{98} $&$ 0 $&$ \frac{31}{16,464} $&$ 0 $&$ -\frac{359}{26,880} $&$ 0 $\\ \hline \hline

$\alpha        $ &$  1.38 $&$  .239 $&$ \dots $&$ .537 $&$ .336 $&$ \dots $&$ \dots $&$ \dots $&$ -.0336 $&$ \dots  $&$ \dots $\\ \hline

$\tilde\alpha $ &$      0  $&$       0 $&$ .194 $&$       0 $&$   0   $&$  -.0918  $&$ .0381  $&$ -.00188 $&$    0    $&$ -.0134 $&$ .0151 $\\ \hline

\end{tabular}
\caption{In the table we report the coefficients of the finite terms in the one-loop corrections to the power spectrum, computed in perturbation theory with dimensional regularization. We omit those cases in which $\alpha$ is degenerate with a fitting parameter. We attach to this publication a Mathematica notebook \cite{file} that generates these numbers and allows to compute their analog for any $n$.}
\end{table}
\end{centering}

The last point to discuss concerns logarithms. For some $n$'s, the loop integrals have a logarithmic divergence, or equivalently a term $\ln (\Lambda/k)$. As customary in field theory, the $\ln \Lambda$ is canceled by a couterterm, but the $\ln k$ remains in the final result. These log terms at one loop come with the same $k$ power law as the finite terms, i.e.~$(k/k_{NL})^{2(3+n)}$, as expected by dimensional analysis. This can also be seen by looking at the scale dependence of the most divergent terms in \eqref{P13div} and \eqref{P22div}, i.e.~$k^{5+n} \Lambda^{n+1}$ and $k^{7}\Lambda^{2n-1}$ (we added a factor of $k^{3}$ as appropriate for discussing $\Delta^{2}$ rather than $P$). When the subleading terms scaling as positive powers of $(k/\Lambda)$ are included, $\Lambda$ becomes $k$ and one obtains again the scaling claimed above. This means that for the counterterms to be able to cancel the $\ln \Lambda$ in the log divergences, one needs $\Delta_{\cs}^{2}$ or $\Delta_{noise}^{2}$ or their higher derivative corrections to have exactly the same $k$-dependence as the loop corrections. For the one-loop corrections considered here, this happens when $2(n+3)=5+n+2m$ or $2(n+3)=7+2m$ for some positive integer $m$, since higher derivative corrections come with an even number of additional powers of $k$ \cite{BNSZ}. Hence we expect one-loop log divergences for $n=-1+2m$ and $n=1/2+m$, with $m$ a positive integer, from $P_{13}$ and $P_{22}$ respectively. This can be confirmed by direct computation\footnote{Notice that using the formulae in appendix B of \cite{FS} one finds a log divergence in $P_{22}$ for $n=1$, which would be in contradiction with the discussion above. The reason is that there is a typo in \cite{FS} in the sign of the fourth term in $P_{22}$, which should be $-11/4$ instead of plus. We re-derive this result in appendix \ref{a:dimreg}. After correcting the typo the log divergent $n$'s from dimensional regularization perfectly match the expectations from the cutoff regularization discussed here.\label{ft}} \cite{file} and can be seen in table \ref{t}, where $\tilde \alpha_{13,22}$ are non-vanishing exactly for $n=-1+2m$ and $n=1/2+m$, respectively\footnote{One subtlety arises for $n=-1$ (see also footnote \ref{IRdiv}), when there are logarithmic IR divergences, which contribute to $P_{13,22}$ computed in dimensional regularization as discussed in appendix \ref{a:FS}. For this case one can check that $\tilde \alpha_{22}=1/3=-\tilde \alpha_{13}$ are exactly the expected IR logarithmic divergences, e.g.~by computing the loop integral with an IR cutoff. E.g.~one can see this in table 5 of \cite{BCGS}, keeping in mind that our numbers are a factor of $(2\pi)^{3}/(2\pi^{2} \pi)=4$ lower than theirs. The $(2\pi)^{3}$ comes from the different Fourier conventions, the factor $2\pi^{2}$ from the fact that we present our results with respect to $k_{NL}$ and finally the last $\pi$ is because of the normalization chosen in that table $P/(\pi A^{2} a^{4})$. In table \ref{t} we have subtracted these log-divergences from the values computed in dimensional regularization $\tilde \alpha_{13}=-44/315$ and $\tilde \alpha_{22}=1/3$} . 

So, after subtracting the divergences with appropriate counterterms, the loop corrections take the form
\be\label{ne}
\Delta^{2}_{1-loop}&=& \left(\frac{k}{k_{NL}}\right)^{3+n} +\left(\frac{k}{k_{NL}}\right)^{2(3+n)} \left[\alpha(n) +\tilde \alpha(n)\ln \left(\frac{k}{k_{NL}}\right)\right]\,.
\ee
Here $\alpha$ and $\tilde  \alpha$ can be computed directly and are given in table \ref{t}. The direct computation using \eqref{P13} and \eqref{P22} is in principle straightforward but rather lengthy, especially for $P_{22}$. In the next subsection we will show how, using the results of \cite{FS}, the $\alpha$'s can be easily computed. \eqref{ne} was given already in \cite{FS} for $n<-1$. The result of our approach shows two qualitative differences. First there are additional corrections to the power spectrum, that, summing up \eqref{ne}, \eqref{csca} and \eqref{jsca},  takes the form
\be
\dsq&=& \left(\frac{k}{k_{NL}}\right)^{3+n} \left\{1+\left(\frac{k}{k_{NL}}\right)^{3+n} \left[\alpha(n) +\tilde \alpha(n)\ln \left(\frac{k}{k_{NL}}\right)\right]\right\}\label{finres}\\
&&\quad +\beta(n) \left(\frac{k}{k_{NL}}\right)^{5+n} \left[1+{\rm higher\,deriv.}\right] +\gamma(n) \left(\frac{k}{k_{NL}}\right)^{7} \left[1+{\rm higher\,deriv.}\right]\,.\nonumber
\ee
Second this result is valid for any $n>-3$ (below which IR-divergences become problematic) and not just for $-3<n<-1$. No divergences appear in the final result and self-similarity is preserved for any $n>-3$.

Before concluding, let us remark that the $\ln k_{NL}$ term in \eqref{ne} comes from $\Delta_{\cs}^{2}$ or $\Delta_{noise}^{2}$ when choosing a particular time dependence (since $k_{NL}$ is a function of time) of $\cs$ and $\cv$ or $\Delta_{J}^{2}$. As anticipated, this time dependence is different from the one we deduced in subsection \ref{sss} using self-similarity. The reason is again that since we are using a regularization that violates self-similarity, so do the counterterms.   Notice that the final result \ref{ne} is self-similar as it should.


\subsection{Dimensional regularization}

Dimensional regularization applied to SPT was first discussed in \cite{FS}. There, formulae were provided for the finite and log-divergent one-loop corrections in three dimensions, since the power law divergences disappear in dimensional regularization. We re-derive those results in any dimension (correcting a typo in $P_{22}$, see footnote \ref{ft}) in appendix \ref{a:FS} and we have included them in an attached mathematica file \cite{file} for the convenience of the reader. One advantage of this regularization procedure is that it respects self-similarity, so that all terms obey the scalings that we computed in section \ref{sss}. Using the dimensionally regularized one-loop integrals $P_{22,13}$ in appendix \ref{a:dimreg}, we have computed the numerical coefficients of the finite one-loop corrections for the $n$'s shown in table \ref{t}. Notice that for each $n$ only one such coefficient is needed in the final formula, since when the log appears, i.e.~when $\tilde \alpha\neq0$, then some fitting parameter, either $\beta$ or $\gamma$, is degenerate with $\alpha$, whose precise value we omit since it is irrelevant for the final shape of the power spectrum. 

Logarithmic divergences arise for those $n$'s for which \eqref{FS1} and \eqref{FS2} are divergent. This happens for $n=-1+2m$ and $n=1/2+m$, with $m$ a positive integer, in agreement with what we found in the previous subsection (see footnote \ref{ft}). In these cases, one needs to expand the $d$-dimensional version of these formulae around the dimension $d=3$ by substituting $d\rightarrow d+\epsilon$ and sending $\epsilon$ to zero. Unfortunately \cite{FS} gives only the results for $d=3$, rather than as function of $d$. We re-derived the general formulae in appendix \ref{a:dimreg}, the result is given in \eqref{P22d} and \eqref{P13d}. Schematically, the leading order of a typical term in \eqref{P22d} or \eqref{P13d} looks like 
\be\label{pom}
P^{2}_{22,13}(k,\tau)\sim\frac{\Gamma (\dots)\Gamma (\dots)\Gamma(\dots)}{\Gamma (\dots)\Gamma(\dots) \Gamma(\dots)} \left(\frac{k}{k_{NL}}\right)^{2n+3+\epsilon}\,,
\ee
where the argument of the $\Gamma$'s depend on $d$ and $n$ (for $P_{13}$ there are only two $\Gamma$'s but this is irrelevant for the following argument). The fact that this expression can be written as function of $k/k_{NL}$ for any $d$ (and $n$) is a consequence of self-similarity, which, as we check in appendix \ref{d}, holds in any number of dimensions (with an appropriate $d$- and $n$-dependent definition of $k_{NL}$ as in \eqref{kNLd}).

Logarithms arise when the argument of the $\Gamma$'s in the numerator (since $\Gamma$ never vanishes, no divergences come from the denominator) become a null or negative integer as $d\rightarrow 3$. Then the expression has a pole $1/\epsilon$ with $\epsilon\equiv d-3$. Using 
\be
\Gamma[-m+\epsilon]=\frac{(-1)^{m}}{m!} \frac{1}{\epsilon}+\mathcal{O}(\epsilon)
\ee
with $m$ a non-negative integer, one can expand \eqref{pom} as
\be\label{fint}
\frac{\#}{\epsilon} \times 2 \epsilon \ln \left(\frac{k}{k_{NL}}\right)+{\rm finite \, terms}\,,
\ee
where $\#$ are the specific numerical factors which determine the precise $n$-dependent value of $\tilde \alpha$ (see table \ref{t}). Notice that the finite terms in \eqref{fint} (determining the coefficient $\alpha$ of the finite loop contribution) are always degenerate with some fitting parameter ($\beta$ or $\gamma$ in \eqref{finres}) and are therefore irrelevant for the shape of the power spectrum. This is a consequence of the fact that log-divergences arise only when the loop corrections have the same $k$ dependence (hence the degeneracy) as the noise or speed of sound corrections, as explained in the previous subsection. Notice that, since the $\ln k_{NL}$ terms come already from the one-loop integrals, the effective coefficients in the terms $\Delta_{\cs}^{2}$ and $\Delta_{noise}^{2}$ now have precisely the time dependence that we derived in subsection \ref{sss} using self-similarity. This was to be expected since dimensional regularization respect the scaling symmetry (see appendix \ref{d}).


\section{Comparison with numerical results}\label{s:num}

In this section, we compare the final predictions of the renormalized EFToLSS with numerical simulations, for self-similar initial conditions in EdS, \eqref{finres}. 

The first point to make is that the EFToLSS makes a definite prediction for any self-similar initial power spectrum $P_{in}(k)=  A k^{n}$ with \textit{any} $n$, as one expects from a self-consistent theory. On the contrary, SPT provides a prediction only for $-3<n<-1$, which highlights some inconsistency in the theory. This is a key point, independent of any phenomenological discussion, so let us stress it again. The EFToLSS, as opposed to SPT, is the theoretically consistend way to study cosmological perturbation theory because it makes physically meaningful predictions (i.e.~cutoff independent) for any initial condition\footnote{Strictly speaking here we only discuss self-similar initial conditions in EdS, but the fact that the EFToLSS has the right counterterms to cancel loop divergences is general.}. Second, as it is common in effective field theories, the predictions of the EFToLSS have fitting parameters, corresponding to the renormalization conditions that needs to be imposed to select a specific solution of the renormalization group flow. As a consequence, the EFToLSS predictions will always fit the numerical (or observational) power spectra as well or better than SPT. To study whether the improvement is just due to the additional fitting coefficients or it really comes from capturing the relevant physics requires a dedicated analysis that accounts for the error bar in the simulation (observations). In addition, one might not be particularly impressed by fitting a smooth line such as the power spectrum with a polynomial. More convincing evidence would come from a realization by realization comparison and from a direct measurement of the short mode stress tensor and thus the EFToLSS parameters in simulations. This has been done in  \cite{CHS} for the $\Lambda$CDM cosmology. We leave this for future work. Finally, when the coefficients $\beta$ and $\gamma$ are fixed by fitting the power spectrum, there is some arbitrariness in how the fitting is performed. We choose to perform a least $\chi^2$ fit over the range $.01<k/k_{NL}<2$ approximating the error bars in the simulation to be $k$-independent. We chose this particular range according to the following considerations. The lower bound comes from the size of the simulations. The upper bound should be of order one, but there is some ambiguity in how the non-linear scale is defined. We used that $\Delta^2(k_{NL})=1$ but other choices are possible, which differ from this by some order-one factor. Operationally, one would like to define $k_{NL}$ by computing higher and higher order corrections and looking for the $k/k_{NL}$ beyond which the fitting cannot be improved. As a rough estimate of this point we take $k/k_{NL}=2$.

\begin{figure}
\centering
\includegraphics[width=\textwidth]{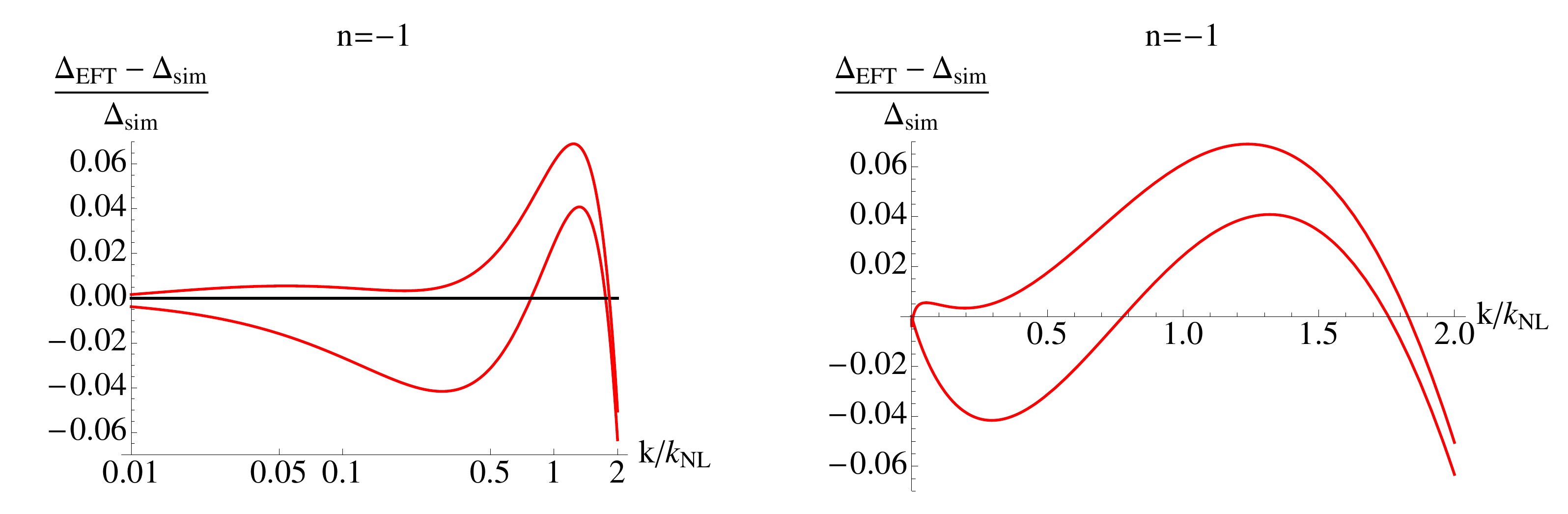}
\includegraphics[width=\textwidth]{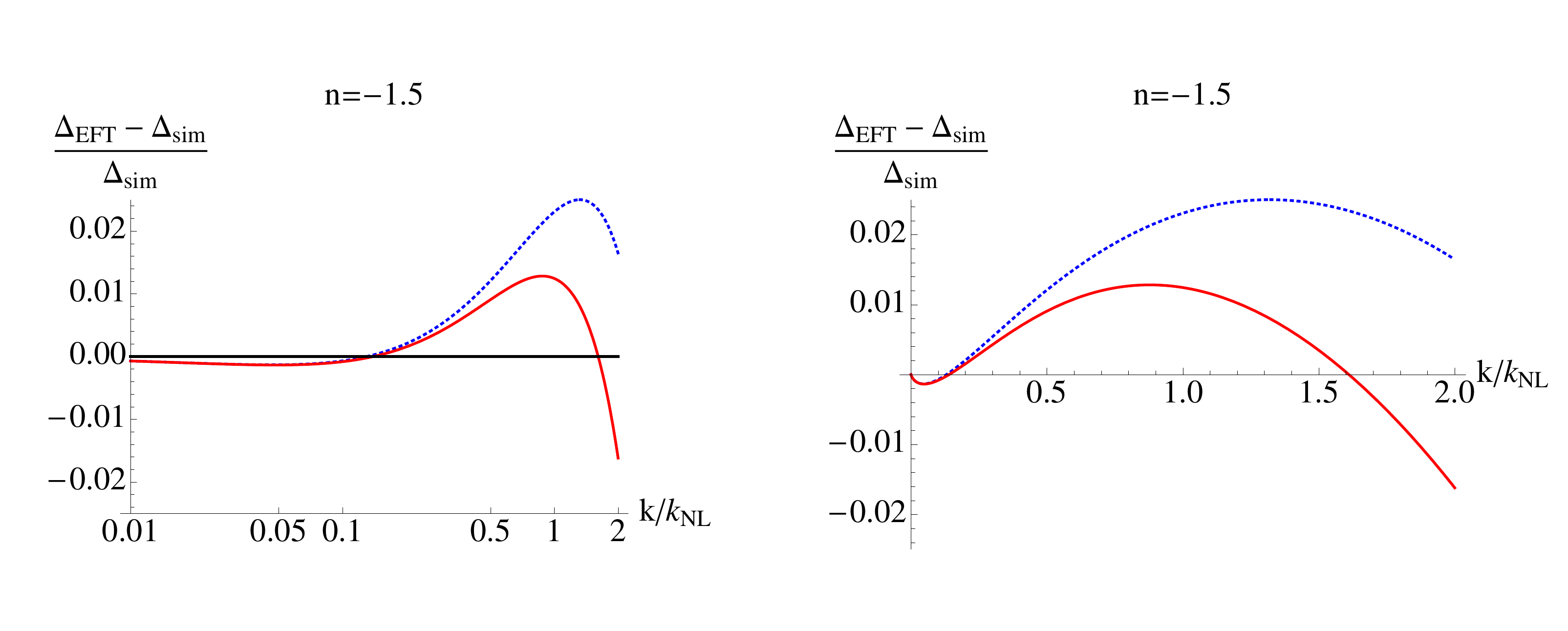}
\caption{The plots show the fractional difference of EFToLSS (red, thick line) and, when available, SPT (blue, dashed line) predictions from the simulations of \cite{OW} for $n=-1.5,-1$ and of \cite{W} for $n=-1$. We show the same plot both with a linear and with a log abscissa. For $ n=-1 $ when both simulations are available the top line refers to \cite{W}. The $k$-range shown and the ambiguity $\mathcal{O} (1)$ in the definition of $k_{NL}$ are discussed in the middle of section 4.\label{f:pt}}
\end{figure}

Now that we have made these cautionary remarks, we can move on and discuss how \eqref{finres} compares with simulations. We use the fitting formulae given in equations A2 of \cite{OW} (for $n=-1.5,-1$) and in 22 of \cite{W} (for $n=-1$). It would be nice in the future to consider other cases as well. For each $n$, we plot in figure \ref{f:pt} the fractional difference of EFToLSS (red, thick line) and, when available, SPT (blue, dashed line) predictions from the fitting formulae extracted from the simulations. In the left and right panels we show the same functions but with a linear or log abscissa, respectively. As shown in figure \ref{ns}, for $-3<n<-1$ the leading correction to the linear power spectrum comes from the one loop term. All of these are known and summarized in table \ref{t}. For $-2<n<-1$ the next-to-leading comes from $\Delta^{2}_{\cs}$. For $n=-2$ the  $\Delta^{2}_{\cs}$ correction is of the same order as the two-loop term, and, as discussed in the previous section, we would expect a log correction to arise (at two loops, i.e.~multiplied by $(k/k_{NL})^{3(3+n)=3}$). Since the coefficient of the finite two-loop term are not known and hard to compute we cannot show plots for this case, even though the results from simulations would be available\footnote{A preliminary look at the $n=-2$ plots, neglecting the unknown two-loop log term, shows that they are qualitatively similar to the $n=-1.5$, which we show in figure \ref{f:pt}.} \cite{W}. Instead for $n>-2$ the two-loop correction is at least the next-to-next-to-leading order and therefore negligible. Notice that $n\sim-1.5$ is the relevant index for $\Lambda$CDM around the non-linear scale, so we expect that for our universe the next-to-leading correction after the one-loop term is $\Delta^{2}_{\cs}$ and \textit{not} the two-loop correction as SPT would suggest. This should be checked directly by computing the EFToLSS parameters directly from simulations of our cosmology.

Let us move on and consider $n=-1.5$ and $-1$. In both cases there is one parameter, namely $\beta$, which we fit as discussed above\footnote{Notice that the parameter $\gamma$ comes from the noise term $\Delta_{noise}^{2}$. For the considered values of $n$, this is expected to be subleading to the two-loop correction, which we do not know. Hence it would be inconsistent to include it for $n\leq -2/3$.}. In the top of figure \ref{f:pt}, we show $n=-1$, for which the SPT prediction is cutoff dependent and hence unphysical. For this value of $n$ both \cite{OW} and \cite{W} provide fitting formulae so we show the fractional difference from both (the top red line refers to \cite{W}). The EFToLSS provides a good fit in the dispalyed range. In the bottom panel, for $n=-1.5$ we compare with the simulations of \cite{OW}. In this case SPT is finite and so we show its prediction as a dashed blue line. One can see that the difference between SPT and EFToLSS increases as one approaches the non-linear scale as expected since the speed of sound and dissipation (and noise) terms grow with some positive power of $k/k_{NL}$. The EFToLSS provides a better fit than SPT over the range shown. The results are not yet conclusive because of the remaining uncertainty in the simulations, the small range of $n$ studied etc. We postpone a more detailed numerical study to a future publication.

 
\section{Discussion}\label{end}

Standard perturbation theory (SPT) is unsatisfactory for at least three reasons. First, it does not have a clear expansion parameter since the density contrast is  \textit{not} small on all scales. Second, it does not consistently account for deviations at large scales from a perfect pressureless fluid induced by short-scale non-linearities. Third, for generic initial conditions loop corrections are UV divergent and therefore the predictions beyond linear theory are cutoff dependent and hence unphysical. 

The recently developed Effective Field Theory of Large Scale Structure (EFToLSS) \cite{BNSZ,CHS,H} successfully addresses all of the above issues. The main idea is to smooth every field (e.g.~density and velocity or momentum) by integrating out the short-scale degrees of freedom beyond a certain cutoff scale. If the cutoff scale is chosen to be larger than the non-linear scale, then the smoothed fields are small everywhere and provide a suitable expansion parameter for perturbation theory. The effect of integrating out the short-scale modes (i.e.~substituting them with the solution of their equations of motion, in an explicit or implicit way) is to induce additional terms in the fluid equations for the large scales. At lowest order these are an effective pressure, shear and bulk viscosity and a stochastic noise, which shows that on large scales the behavior of collisionless dark matter deviates from that of a pressureless perfect fluid\footnote{These corrections, analogously to the non-linear corrections, go to zero as $k\rightarrow 0$. Neveetheless, if one wants to make accurate predictions, it is important to consistently include these corrections together with the non-linear ones, with the specific details depending on the initial power spectrum, as discussed in this work.}.

In this paper we have focused on the third issue mentioned above, i.e.~the fact that SPT prediction are cutoff dependent and hence unphysical for a wide range of initial conditions. In the case of our universe, according to the $\Lambda$CDM model, these UV divergences are absent because the initial power spectrum is sufficiently red-tilted around the non-linear scale, with approximately $n\sim -1.5$. But this notwithstanding, one would like to describe our universe using a well-understood theory that is consistent with generic initial conditions and not just for specific ones. As our discussion here shows, investigating the theoretical consistency of the theory, even for phenomenologically non-viable initial conditions, highlights that we are missing some ingredient in the construction. In the present case, these ingredients are the effective coefficients induced by integrating out the short-scale. As pointed out before \cite{BNSZ,CHS,H}, and discussed here in more details, these terms (speed of sound, viscosity and stochastic noise) have precisely the right $k$-dependence to cancel the divergences arising in perturbation theory beyond linear order. This implies that the EFToLSS, rather than SPT, is the consistent way to study the cosmological perturbation theory of structures. 

After we have convinced ourselves that these additional effective terms beyond a perfect fluid are needed for the consistency of the theory we can start asking about the role that they play in phenomenological predictions. In order to assess the relative importance of various corrections we have studied here self-similar solutions in EdS. These are easy to handle analytically for a wide range of initial conditions and are not too far from realistic models. As we summarize in figure \ref{ns}, depending on the index $n$ in the self-similar initial power spectrum $P_{in}=A  k^{n}$, the relative importance of the various corrections changes. For $-3<n<-2$ one- and two-loop corrections are the leading and next-to-leading order corrections\footnote{For $n<-3$ the theory is plagued by IR divergences.}. For $-2<n<-1$ one-loop is the leading correction and the speed of sound the next-to-leading one. This interval is expected to be the most relevant for phenomenology since in our universe around the non-linear scale $n\sim-1.5$. This is an important result since it shows that, for our universe, the corrections present in the EFToLSS, but neglected in SPT, are more important than two-loop corrections. A detailed study of the EFToLSS description of a realistic $\Lambda$CDM model was carried out in \cite{CHS} (see also \cite{H}). Here we have also compared the predictions of the EFToLSS with existing simulation of self-similar solution in EdS, in particular for $n=-1.5$ and $-1$. They fit the data well, but further analysis will be able to provide much more convincing evidence as we discuss below. As one goes towards larger $n$ the EFToLSS corrections become even more important than the one-loop corrections.

There are many interesting possibilities for future research. It would be appropriate to perform dedicated simulations of self-similar solutions in EdS to compare with the EFToLSS. Ideally, this would extend our results here towards larger $n$'s for which the relative importance of EFToLSS increases. Also, one would like to focus on large scales where any subleading correction is very small, rather than close to the non-linear scale where all corrections tend to be of the same order. Other statistics beside the density power spectrum should be considered, and in particular comparisons performed realization by realization along the lines of \cite{TS123}. One could study higher orders of perturbation theory with the goal of better defining the non-linear scale. 

Finally, in recent years it has become apparent that, at least for parameters close to those of our Universe, Lagrangian perturbation theory is a better starting point for a perturbative expansion \cite{TS123}. The EFT approach has so far only been developed in the Eulerian context. It would be good to understand how the EFT ideas could be implemented in a Lagrangian scheme and thus perhaps improve it even further. 



 
\section*{Acknowledgements}

We would like to thank R.~Scoccimarro, L.~Senatore and S.~Tassev for useful discussions and comments on the draft. E.~P.~is supported in part by the Department of Energy grant DE-FG02-91ER-40671. M.~Z.~is supported in part by the National Science Foundation grants PHY-0855425,  AST-0907969,  PHY-1213563 and by the David \& Lucile Packard.


\appendix


\section{Details of the dimensional regularization}\label{a:FS}

In \cite{FS} the results for the dimensional regularization of the $P_{22}$ and $P_{13}$ integrals were given for $d=3$. From those formulae one can easily extract all the finite terms in the loop integral for those $n$ for which there is no log-divergence. For the $n$'s leading to a log-divergence in three-dimensions, we need to know how the loop integrals depend on $d$ explicitely, in order to be able to take the limit $\epsilon \rightarrow 0$ where $\epsilon\equiv d-3$ and extract the $\epsilon^{-1}$ pole as well as the finite term $\epsilon^{0}$. Hence in the following we re-compute the loop integrals in dimensional regularization and give and example of how the coefficient of the log-divergence is computed.

 
\subsection{One-loop integrals}\label{a:dimreg}

In this section we re-derive the one-loop integrals following \cite{FS}, but giving explicit formulae for the $d$ dependence of the final result. For this computation we need the following formulae for the power spectrum (see e.g.~\cite{FS}) with self-similar initial conditions $P_{in}(k)=A k^{n}$
\be
P_{22}(k,\tau)  &=&   A^2 a^4 (\tau)  \int \frac{d^d {\bf q}}{(2\pi)^{3}} \ q^n2 |{\bf k}-{\bf q}|^n \,[F_2^{(s)}({\bf k}-{\bf q},{\bf q}) ]^2   \,,\\
         P_{13}(k,\tau )  &=&   A^2 a^4 (\tau)  
\int \frac{d^d {\bf q}}{(2\pi)^{3}} \ q^n  6  k^n   F_3^{(s)}({\bf k},{\bf q},-{\bf q}) \,,
\ee
with the kernels given by
\be
F_2^{(s)}({\bf q}_1,{\bf q}_2) &=& \frac{5}{7} + \frac{1}{2} 
	 \frac{{\bf q}_1 \cdot {\bf q}_2}{q_1 q_2} \left(\frac{q_1}{q_2} +
	 \frac{q_2}{q_1}\right) + \frac{2}{7} \frac{({\bf q}_1 \cdot {\bf 
	 q}_2)^2}{q_1^2 q_2^2}\,,\\
F_3^{(s)}({\bf k},{\bf q},-{\bf q}) &=& \frac{1}{|{\bf k}-{\bf q}|^{2}}\left[\frac{5k^{2}}{126}-\frac{11 {\bf k}\cdot {\bf q}}{108}+7 \frac{({\bf k}\cdot {\bf q})^{2}}{108 k^{2}}-\frac{k^{2} ({\bf k}\cdot {\bf q})^{2} }{54 q^{4}}+\frac{4 ({\bf k}\cdot {\bf q})^{3}}{189 q^{4}}-\frac{23 k^{2} {\bf k}\cdot {\bf q}}{756 q^{2}}\right.\nonumber\\
&&\left.+\frac{25 ({\bf k}\cdot {\bf q})^{2}}{252 q^{2}}  -\frac{2 ({\bf k}\cdot {\bf q})^{3}}{k^{2}q^{2}}\right] +\frac{1}{|{\bf k}+{\bf q}|^{2}}\left[\frac{5 k^{2}}{126} +\frac{11{\bf k}\cdot {\bf q} }{108 }-\frac{7 ({\bf k}\cdot {\bf q})^{2}}{108 k^{2}} \right.\nonumber\\
&&\left. -\frac{4 k^{2} ({\bf k}\cdot {\bf q})^{2} }{27 q^{4}}-\frac{53 ({\bf k}\cdot {\bf q})^{3}}{189 q^{4}} +\frac{23 k^{2} {\bf k}\cdot {\bf q}}{756 q^{2}}-\frac{121 ({\bf k}\cdot {\bf q})^{2}}{756 q^{2}} -\frac{5 ({\bf k}\cdot {\bf q})^{3}}{27 k^{2}q^{2}} \right]\,,
\ee
where, instead of the general $F^{s}_{3}$ (see e.g. \cite{BCGS}), we have written down the expression needed for the one-loop power spectrum. We will also need the mathematical identities
\be
\int  \frac{d^d q}{(q^2)^{\nu_1} [({\vec k}-{\vec q})^2]^{\nu_2}} &=&
 \frac{\Gamma (d/2 -\nu_1)\label{sop}
\Gamma (d/2 -\nu_2)
\Gamma (\nu_1 + \nu_2 - d/2)}{\Gamma (\nu_1) \Gamma (\nu_2) \Gamma (d-
\nu_1 - \nu_2)} \ \pi^{d/2}  \ k^{d-2 \nu_1 -2 \nu_2}\,,\\
{\bf k}\cdot {\bf q}&=&\pm \frac{1}{2}\left(k^{2}+q^{2}-|{\bf k}\mp{\bf  q}|^{2}\right)\,. \label{piu}
\ee
The computation can then be performed using Mathematica or a similar software. We attach to this publication a file \cite{file} that outputs the final result. For $P_{22}$ one can straightforwardly use \eqref{piu} with the upper sign to get $P_{22}$ in the form of a sum of terms like those appearing in \eqref{sop}. For $P_{13}$ one needs to use \eqref{piu} with the lower sign for the terms with a factor $|\vec k+\vec q|^{-2}$, while \eqref{piu} with the upper sign for the others. The final result as function of $d$ and $n$ is
\begin{eqnarray}
        P_{22}(k,\tau) & =&  \Bigg(
\frac{\Gamma [4- \frac{d}{2}-n] \Gamma^2 [(-4+d+n)/2]}{2 \Gamma^2 (2-n/2) \Gamma [-4+d+n]} +
\frac{3 \Gamma [3- \frac{d}{2} -n] \Gamma [(-4+d+n)/2] \Gamma [(-2+d+n)/2]}{\Gamma [1-n/2] \Gamma[2-n/2]\Gamma [-3+d+n]}  
 \nonumber \\ & & 
+\frac{29 \Gamma [2- \frac{d}{2} -n] \Gamma^2 [(-2+d+n)/2]}{4 \Gamma^2[1-n/2] \Gamma [-2+d+n]}
-\frac{11 \Gamma [2- \frac{d}{2} -n] \Gamma [(-4+d+n)/2] \Gamma [(d+n)/2]}{
4 \Gamma [2-n/2] \Gamma [-n/2] \Gamma [-2+d+n]} \nonumber \\ & & - 
 \frac{15 \Gamma [1- \frac{d}{2} -n] \Gamma [(-4+d+n)/2] \Gamma [(2+d+n)/2]}{
 2 \Gamma [-1-n/2] \Gamma [2 -n/2] \Gamma [-1+d+n]}
+ \frac{15 \Gamma [1- \frac{d}{2} -n] \Gamma [(-2+ d+n)/2]}{
2 \Gamma [1-n/2] \Gamma [-n/2]} \nonumber \\ & &
\times \frac{ \Gamma [(d+n)/2]}{ \Gamma (-1+d+n)} 
-\frac{25 \Gamma [-d/2 -n] \Gamma [(-2+d+n)/2] \Gamma [(2+d+n)/2]}{
 \Gamma [-1-n/2] \Gamma [1-n/2] \Gamma [d+n]} 
 + \frac{25 \Gamma [-d/2 -n] }{4 \Gamma [-2-n/2]} \nonumber \\ & &
\times \frac{ \Gamma [(-4+d+n)/2] \Gamma [(4+d+n)/2]}{ \Gamma [2-n/2] \Gamma [d+n]}
+  \frac{75 \Gamma [- \frac{d}{2} -n] \Gamma^2 [(d+n)/2]}{4 \Gamma^2[-n/2] \Gamma [d+n]}
 \Bigg) \ \frac{  A^2 a^{4(d-2)}}{49} \frac{1}{8\pi^{3-d/2}} \nonumber \\
& & \times  k^{2n+d}, \label{P22d}\\
         P_{13}(k,\tau) & = & \Gamma[-1+d/2]\Bigg(
- \frac{\Gamma [(4-d-n)/2] \Gamma [(-2+d+n)/2]}{84 \Gamma (1-n/2) \Gamma [-2+d+n/2]}
- \frac{19 \Gamma [-(d+n)/2] \Gamma [(2+d+n)/2] }{84 
 \Gamma (-1-n/2) \Gamma (d+n/2) } \nonumber \\ & & +
\frac{ \Gamma [-(2+d+n)/2] \Gamma [(4+d+n)/2]}{12 \Gamma[-2-n/2] \Gamma [1+d+n/2]}
+ 
\frac{5  \Gamma [(2-d-n)/2] \Gamma [(d+n)/2]}{
28 \Gamma [-1+d+n/2] \Gamma [ -n/2] } \nonumber \\ &  & -
 \frac{  \Gamma [(-4+d+n)/2] \Gamma [(6-d-n)/2]}{
42 \Gamma [2-n/2] \Gamma [-3+d+n/2] }  \Bigg) \ \frac{1}{8\pi^{3-d/2}}  A^2 a^{4(d-2)} \
k^{2n+d}. \label{P13d}
\end{eqnarray}
For $d=3$ these general formulae reduce to
\begin{eqnarray}
        P_{22}(k,\tau ; n) & =&  \Bigg(
\frac{\Gamma (5/2 -n) \Gamma^2 [(n-1)/2]}{2 \Gamma^2 (2-n/2) \Gamma (n-1)} +
\frac{3 \Gamma (3/2 -n) \Gamma [(n-1)/2] \Gamma [(n+1)/2]}{
 \Gamma (1-n/2) \Gamma (2-n/2) \Gamma (n)} \ \ \ \ \ \ \ \ \ \ 
 \nonumber \\ & & +  
\frac{29 \Gamma (1/2 -n) \Gamma^2 [(n+1)/2]}{4 \Gamma^2[1-n/2] \Gamma (n+1)}
-
\frac{11 \Gamma (1/2 -n) \Gamma [(n-1)/2] \Gamma [(n+3)/2]}{
4 \Gamma (2-n/2) \Gamma ( -n/2) \Gamma (n+1)} \nonumber \\ & & - 
 \frac{15 \Gamma (-1/2 -n) \Gamma [(n-1)/2] \Gamma [(n+5)/2]}{
2 \Gamma (-1-n/2) \Gamma (2 -n/2) \Gamma (n+2)} + 
\frac{15 \Gamma (-1/2 -n) \Gamma [(n+1)/2]}{
2 \Gamma (1-n/2) \Gamma (-n/2)} \nonumber \\ & &
\times \frac{ \Gamma [(n+3)/2]}{ \Gamma (n+2)} -  
 \frac{25 \Gamma (-3/2 -n) \Gamma [(n+1)/2] \Gamma [(n+5)/2]}{
 \Gamma (-1-n/2) \Gamma (1-n/2) \Gamma (n+3)} + 
\frac{25 \Gamma (-3/2 -n) }{
4 \Gamma (-2-n/2)} \nonumber \\ & &
\times \frac{ \Gamma [(n-1)/2] \Gamma [(n+7)/2]}{ \Gamma (2-n/2) \Gamma (n+3)}
+  \frac{75 \Gamma (-3/2 -n) \Gamma^2 [(n+3)/2]}{4 \Gamma^2[-n/2] \Gamma (n+3)}
 \Bigg) \ \frac{  A^2 a^4 (\tau)}{49} \frac{1}{8\pi^{3/2}} \nonumber \\
& & \times  k^{2n+3}, \label{FS1}\\
         P_{13}(k,\tau ; n) & = & \Bigg(
- \frac{\Gamma [(n+1)/2] \Gamma [(1-n)/2]}{84 \Gamma (1-n/2) \Gamma (1+n/2)}
- \frac{19 \Gamma [-(n+3)/2] \Gamma [(n+5)/2] }{84 
 \Gamma (-1-n/2) \Gamma (3+n/2) } \nonumber \\ & & +
\frac{ \Gamma [-(n+5)/2] \Gamma [(n+7)/2]}{12 \Gamma[-2-n/2] \Gamma (4+n/2)}
+ 
\frac{5  \Gamma [-(n+1)/2] \Gamma [(n+3)/2]}{
28 \Gamma (2+n/2) \Gamma ( -n/2) } \nonumber \\ &  & -
 \frac{  \Gamma [(n-1)/2] \Gamma [(3-n)/2]}{
42 \Gamma (2-n/2) \Gamma (n/2) }  \Bigg) \ \frac{1}{8\pi}  A^2 a^4 (\tau) \
k^{2n+3}. \label{FS2}
\end{eqnarray}
which is almost identical to the one in \cite{FS} except for an additional factor of $(2\pi)^{-3}$, due to our different conventions for the Fourier transform, and the sign of the fourth term in $P_{22}$, which we find to be minus instead of plus. Only with our sign the log-divergences in $P_{22}$ arise for $n=1/2+m$, with $m$ a positive integer, as expected from direct inspection of the cutoff integrals (see subsection \ref{cutoff}).

 
\section{Self-similarity in EdS in d-dimensions}\label{d}

In this appendix we collect some results about perturbation theory in d-dimensions in EdS ($\Omega_{m}=1$), which are useful when performing dimensional regularization. From the conservation of the matter energy-momentum tensor at zeroth order in perturbations one finds for the average energy density $\bar\rho\, a^{d}={\rm const}$ and from the Friedman equation
\be
\frac{d(d-1)}{2}  \cH^{2}= \kappa(d) a^{2}\bar \rho\,,
\ee
with $\kappa(d)$ the d-dimensional Newton constant, one finds 
\be
\cH\equiv \dot a=\cH_{0} a^{1-d/2}=\frac{1}{(d/2-1)\tau}\,,
\ee
where for convenience we have set $a_{0}=1$ and $\tau_{0}^{-1}=\cH_{0}(d/2-1)$, so that one has $\cH=2/\tau$ in three dimensions. The non-linear system of fluid and Einstein equations before smoothing and neglecting vorticity, is given by \cite{Eng}
\be
\partial_{\tau} \dl+\partial_{i} \left[ \left(1+\dl\right) v_{l}^{i} \right]&=&0\,,\\
\partial_{\tau}v^{i}_{l}+\cH v^{i}_{l}+\partial_{i} \phi +v_{l}^{k}\partial_{k} v_{l}^{i}&=&0\\
\partial_{i}\partial_{i} \phi_l&=& \frac{d-2}{d-1} a^{2} \bar \rho \dl= \frac{d(d-2)}{2} \cH^{2}  \dl\,,
\ee
i.e.~only the Friedman equation is modified. As we did in section \ref{sss}, one can immediately check that given a solution $[\delta(x,\tau), v^{i}(x,\tau),\phi(x,\tau)]$, another one is generated by the scaling
\be
\tilde \delta(x,\tau)&=&\delta(\lambda_x x, \lambda_\tau\tau)\,, \\
\tilde v^i(x,\tau)&=&{\lambda_\tau \over \lambda_x} v^i(\lambda_x x, \lambda_\tau\tau)\,,\\
\tilde \phi(x,\tau)&=&\left({\lambda_\tau \over \lambda_x}\right)^2\phi(\lambda_x x, \lambda_\tau\tau)\,,
\ee 
for any $\lambda_{\tau}$ and $\lambda_{x}$. The new solution corresponds to the same cosmology only if it has the same power spectrum of initial conditions. For this we need to know the growth factor. The linearized equation of motion for the density contrast is then given by 
\be
-a^{2}\cH^{2} \delta_{,aa}-\frac{6-d}{2} a  \cH^{2} \delta_{,a}+\frac{d(d-2)}{2}\cH^{2}\delta&=&0\,, \quad {\rm or}\\
-\delta_{,\tau\tau}-  \cH  \delta_{,\tau}+\frac{d(d-2)}{2}\cH^{2}\delta&=&0\,,
\ee
and therefore the two independent solutions are power laws: $a^{d-2}\sim \tau ^{2}$ and $a^{-d/2}\sim\tau^{-d/(2-d)}$. In passing, the Green's function is then found to be
\be
G(a,\tilde a)=\theta(a-\tilde a) \frac{2 \, \tilde a^{d-3}}{\cH_{0}^{2}(3d-4)} \left[ \left(\frac{a}{\tilde a}\right)^{-d/2}-\left(\frac{a}{\tilde a}\right)^{d-2}\right]\,.
\ee
where $\theta(a-\tilde a)$ is the Heaviside theta function. The linear power spectrum, valid at early times and neglecting the decaying mode, depends on time as $\Delta^{2}(k,\tau)\propto k^{d} P(k,\tau) \propto \tau^{4}$, i.e.~the same in any dimension. The requirement that the rescaled solution represents the same cosmology as the original one is $\tilde \Delta^2(k,\tau)= \Delta^2(k,\tau)$ which, given that $\tilde \Delta^2(k,\tau)= \Delta^2(k/\lambda_x,\lambda_\tau \tau)$, is satisfied if and only if
\be\label{re}
\lambda_x=\lambda_\tau^{4\over n+d}.
\ee
for an initial power spectrum $P_{in}\propto k^{n}$. So we conclude that in d-dimensions $\Delta^{2}(k,\tau)$ has to be a function of $k/k_{NL}$ with
\be\label{kNLd}
k_{NL}^{d+n}\equiv \frac{2 \pi^{2}}{A a^{2(d-2)}}= \frac{2 \pi^{2}}{A } \left(\frac{\tau_{0}}{\tau}\right)^{4},
\ee
so that it is invariant under the rescaling \eqref{re}.


\end{document}